\documentstyle[12pt,aasms4]{article}

\def\ga{\mathrel{\mathpalette\fun >}}

\def\fun#1#2{\lower0.837ex\vbox{\baselineskip0ex\lineskip0.209ex
  \ialign{$\mathsurround=0ex#1\hfil##\hfil$\crcr#2\crcr\sim\crcr}}}

\def\msunyr{M_\odot \ {\rm yr}^{-1}}
\def\sles{\lower2pt\hbox{$\buildrel {\scriptstyle <}
   \over {\scriptstyle\sim}$}}
 
\def\sgreat{\lower2pt\hbox{$\buildrel {\scriptstyle >}
   \over {\scriptstyle\sim}$}}

\def\ga{\mathrel{\mathpalette\fun >}}

\begin{document}

 \title{A Numerical Gamma-Ray Burst Simulation Using Three-Dimensional
  Relativistic Hydrodynamics: 
 The Transition from Spherical to Jet-like Expansion }

 \vskip  0.75truein

 \author{John K. Cannizzo\footnote{
also University of Maryland    Baltimore County}}
 \affil{e-mail: cannizzo@stars.gsfc.nasa.gov}
 \affil{NASA/GSFC/Laboratory for High Energy Astrophysics, 
 Code 661, Greenbelt, MD 20771}
 \authoraddr{NASA/GSFC/Laboratory for High Energy Astrophysics, 
 Code 661, Greenbelt, MD 20771}

\medskip
\medskip

 \author{ Neil Gehrels }
 \affil{e-mail: gehrels@lheapop.gsfc.nasa.gov}
 \affil{NASA/GSFC/Laboratory for High Energy Astrophysics, 
 Code 661, Greenbelt, MD 20771}
 \authoraddr{NASA/GSFC/Laboratory for High Energy Astrophysics, 
 Code 661, Greenbelt, MD 20771}

\medskip
\medskip

 \author{Ethan T. Vishniac}
 \affil{e-mail: ethan@pha.jhu.edu}
 \affil{Department of Physics and Astronomy,
    Johns Hopkins University,
   3400 N. Charles Street,
   Baltimore, MD 21210} 
 \authoraddr{Department of Physics and Astronomy,
    Johns Hopkins University,
   3400 N. Charles Street,
   Baltimore, MD 21210}

\medskip
\medskip
\medskip
\medskip
\medskip
\medskip
\medskip
\medskip
\medskip
\medskip
\medskip
\medskip
\medskip
\medskip
\medskip
\medskip


\centerline {\bf to appear in the Astrophysical Journal}

\received{  2003 June 9}
\accepted{  2003 October 3}

\begin{abstract}

We present the first
  unrestricted,
 three-dimensional relativistic hydrodynamical
calculations of the blob of gas associated
   with the jet producing  a gamma-ray burst.
   We investigate the deceleration phase
  of the blob corresponding 
      to the time when
  afterglow radiation is produced,
   concentrating on
the transition in which the
   relativistic beaming 
 $\gamma^{-1} $ goes from being less than  $\theta$, where
 $\gamma$ is the bulk Lorentz factor and 
     $\theta$
is the angular width of the jet, to  $\gamma^{-1}$ greater than $\theta$.
      We study the time dependent evolution of
  the physical parameters associated with the jet,
     both parallel to the direction of motion
  and perpendicular to it. 
    We  
    calculate  light curves
    for observers at varying
    angles with respect to the velocity vector
  of the blob, assuming
  optically thin emission
  that scales with the local pressure.
   Our main findings are that
 (i)   gas ahead of the advancing blob  does not
  accrete onto  and  merge with the blob material
   but rather flows around the blob, 
 (ii)  the decay light curve
  steepens
     at a time corresponding roughly to
 $\gamma^{-1} \approx \theta$ 
          (in accord with earlier studies),
   and
(iii) the rate of decrease of the
  forward component
  of  momentum in the blob  is well-fit by a simple
  model in which the
    gas in front of the blob exerts 
   a drag force on the blob, and 
   the  cross sectional area of the blob increases
     quadratically with laboratory time (or distance).

\end{abstract}

\medskip
\medskip

{\it  Subject headings:}
gamma rays: bursts - hydrodynamics - relativity - shock waves

\section{ Introduction }

   Gamma-ray bursts are the most powerful explosions
in the Universe. 
     (For reviews see Piran 1999 and  M\'esz\'aros 2002.)
  A crucial advance
    in  understanding gamma-ray bursts
   began with the discovery of ``afterglows'',
  starting with
      {\it BeppoSax} observations in the soft X-ray band  of GRB 970228
  (Costa et al. 1997; Wijers, Rees, \& M\'esz\'aros 1997).
 (For a review of GRB afterglows see
   van Paradijs, Kouveliotou, \& Wijers 2000.)
    If  GRBs were isotropic, then the measured  redshifts
 would imply
total explosion energies of
            $\sim10^{52} - 10^{54}$ ergs (Frail et al. 2001).
 Theoretical work on relativistic jet expansion, however,
                  shows that one expects
 a steepening  in the  decay light curve if one is looking down
the axis of a jet as the flow decelerates from 
 a bulk  Lorentz factor 
      $\gamma^{-1} <  \theta$ to  $\gamma^{-1}  > \theta$,
where $\theta$ is the  jet beaming angle (e.g., Rhoads 1997,
Sari, Piran, \& Halpern 1999, Panaitescu \& M\'esz\'aros 1999,
   Panaitescu \& Kumar 2001ab, 2002).
      One does in fact see such 
steepenings in the light curves (e.g., Stanek et al. 1999;
                                       Harrison et al. 1999).
    Prior to the time when $\gamma^{-1}\simeq \theta$ 
    the expansion is
effectively ``spherical''
     from the observer's viewpoint because
the relativistic beaming is narrower
 than the jet itself.
   In other words, if the GRB emission
   were coming from one spot 
   on  a large, relativistically expanding sphere, 
  aimed directly at the observer, 
   the
  observer would not see any emission from the
other parts of the sphere.
      After the time when  $\gamma^{-1}\simeq \theta$ 
    the observer
  can ``see'' the entire jet, and
 a faster rate of decline in the luminosity is predicted.
    A separate issue
      that we will address in this work is the sideways 
   or lateral
    expansion of the jet as  the increasing solid angle of the jet 
    enables a larger fraction of the circumstellar medium (CSM)
 surrounding the progenitor star
    to be intercepted and provide
    decelerating gas.
   Several groups have claimed that this 
      leads to a faster (exponential)
      decrease in $\gamma$, which
acts as an additional agent to  diminish
    the amplitude of the relativistic beaming.

   The concept of a ``break'' in the afterglow light
   curves occurring when $\gamma^{-1}\simeq \theta$
   has been used to infer
   the presence of strong beaming in GRBs 
      (Frail et al. 2001; Panaitescu \& Kumar 2001ab, 2002;
                          Panaitescu, M\'esz\'aros, \& Rees 1998;
            Piran, Kumar, Panaitescu, \& Piro 2001). 
 Frail et al. (2001) utilize the theoretical framework 
  of Sari, Piran, \& Halpern (1999) 
   which takes the jet evolution to be spherical adiabatic
   expansion   
 to show that,
after correcting the ``isotropic'' energies to account for
the specific beaming factor for each burst, the total burst
energy reduces to a narrow range centered on $\sim 5\times 10^{50}$ ergs.
  Frail et al. (2001) give a tabulation of 17 afterglows for
which redshifts had been measured, up to 2001 January. The
redshifts range from 0.433 to 4.5. Of these, 10 also have
  break times known to within $\sim30$\%,
           ranging from 1 d to 25 d.
  (In addition, three GRBs have listed lower limits  on the
break time, and two have listed upper limits.)
  The combination of redshifts, fluences, and break times lead to
estimates of the  jet angles $\theta$ ranging from  
 $1^{\circ}$ to $25^{\circ}$, with an aggregation near $4^{\circ}$.
  (The number of GRB's with redshift determinations is
  currently $\sim40$.
     For an update of the work described in Frail et al. 2001,
   see Berger,  Kulkarni, \& Frail 2003.)

Prior  investigations of GRB jet expansion have been largely
analytical,
  or involved computational
  models with some imposed 
   symmetry, typically spherical or axial  
   (e.g., Rhoads 1997, 1999; Sari 1997,
Sari, Piran, \& Halpern 1999; Panaitescu \& M\'esz\'aros 1999;
  Granot, Piran, \& Sari 1999a;
  Kumar \& Panaitescu 2000ab; 
    Panaitescu \& Kumar 2001ab, 2002;  
   Granot et al. 2002).
 These works
      have tended for the most part  to confirm  estimates and
scalings based on the analytical formalism of Blandford \& McKee (1976=BM76),
 although for many of the studies the agreement
   is (to some extent) circular, given that
  they  rely at least in part on the BM76  formalism.
 Kobayashi, Piran, 
    \& Sari (1999)  present results based
on a spherically symmetric relativistic Lagrangian code. They
identify three regimes during the evolution of the GRB jet:
     (i) an acceleration
phase during which $p >> \rho$ and $\gamma$ increases rapidly to a large
value $\sim10^2  - 10^4$, (ii) a coasting phase  during which $\gamma$ is
relatively constant as the  mass accumulated from the CSM is small
compared to that in the  jet, and (iii) a deceleration phase during
which the accumulated mass forces a rapid decrease in $\gamma$.
    Previous studies using analytical methods  divided 
the       evolution into regimes defined by some ordering of
distance, velocity, or energy scales (see Table 3 of Piran 1999  and
                                        Figure 7  of  M\'esz\'aros 2002).  
           The  main findings  that seem to be
common to all studies are that the break in the decaying light
curve
 $ d \log L(t)/d \log t$ occurs  roughly when the deceleration
has decreased $\gamma$ to  roughly the reciprocal of the jet beaming
angle, and that the subsequent decrease in $\gamma$ is roughly
exponential with distance.
  Also,
    Panaitescu \& M\'esz\'aros (1999) present axisymmetric
  calculations to study
   the combined effects of
   the transition from $\gamma^{-1} <  \theta$ to  $\gamma^{-1}  > \theta$
 and the lateral jet expansion.

 In 
other subdisciplines of astronomy  the use of relativistic
  hydrodynamics  codes has been standard for some time.
   For example, workers studying extragalactic jets have used
  such codes to continuously inject a collimated supersonic beam 
   into
  a surrounding medium, usually under the assumption of
  pressure equilibrium.
  (see,
     e.g., Norman et al. 1982 for a thorough discussion).
 In the context of GRBs, 
   work has been done using 2D and 3D relativistic 
                    hydro codes to consider the
   evolution of the GRB jet as it propagates through the
   envelope of the progenitor star, up to the point where
  it breaks out of the stellar surface 
 and produces the prompt GRB emission
(Zhang, Woosley, \& MacFadyen 2003,
 Zhang, Woosley, \& Heger     2003).
   In this 
    work we consider the evolution
  covering the afterglow  time (i.e, after the period
  considered by Zhang et al. 2003ab). 
       We utilize a  three dimensional relativistic
hydrodynamical code to study the propagation of an initially ultrarelativistic
blob into a dense CSM. We study the spatial spreading of the
blob both along the direction of propagation and orthogonal to it,
as well as the evolution of  $\gamma$ in space and time. 
   We also calculate afterglow light curves, taking a simple
prescription in which the local emissivity scales with the local
  pressure.

\section { Computational Model and Tests }


 The  model used is that of Del Zanna \& Bucciantini (2002=DB02).
  These authors present a simple and efficient  numerical
scheme for special relativistic hydrodynamics (SRHD)  that does not
rely on computationally expensive spectral decomposition (and 
the accompanying matrix inversions). The semi-discrete form
 of the SRHD equations is solved, so that time integration
 can be carried out using a standard Runge-Kutta method.
      Unlike many previous multidimensional calculations run
on supercomputers, characteristic decomposition and Riemann solvers
   are not required. The only local pieces of information
required are the highest characteristic speeds.
  The simplicity and efficiency of the model allow for runs to be
  carried out  on a modern PC with $\sim2$ GHz
CPU speed and $\sim1$ Gb  RAM.
   We 
refer the reader to DB02
      and  to Londrillo \& Del Zanna (2000)  for  the details.
 (Also, 
   Del Zanna \& Bucciantini 2003 add 
    MHD to the SRHD formalism of DB02.)
 The
   basic idea is to calculate a vector of  
               ``fluxes'' {\bf f} that is then used
  within each time step  to
    advance a vector of ``conserved'' variables {\bf u} 
   by computing the spatial divergences of the {\bf f}'s.
 The time integration is carried out with  a  third order
 Runge-Kutta scheme.
 The vector of ``primitive'' or physical variables
  ${\bf v} = [\rho, \ v^j, \ p]^T$ must be recovered
  from the conserved
 variables ${{\bf u}({\bf v})} = [\rho\gamma, \ w\gamma^2 v^j, \ w\gamma^2 - p]^T$
in each Runge-Kutta step,
  and at each grid point. 
 Primitive 
        variables are reconstructed at the left (L) and right (R)
cell faces using the ``convex essentially non-oscillatory''
interpolation described in DB02. This method gives third order
accuracy by  adaptively adjusting the ``stencil'', or spatial interval,
used for computing differences in the vicinity  of a given cell.
  DB02 
     present two versions of the fluxes, 
 {\bf f}$^{\rm HLL}$,
 where ``HLL'' denotes  Harten, Lax, \& Van Leer (1983), 
   that are precise  but at the same time
  potentially  prone to numerical instabilities, and
{\bf f}$^{\rm LLF}$,
  where ``LLF'' denotes local Lax-Friedrichs (cf. Lax \& Liu 1998),
  that represent a
 smoother, numerically dissipative flux. 
  For the main application we consider in this work 
 we will utilize the LLF fluxes.

   DB02 present a suite of test results for 1, 2, and 3 dimensions.
We have reproduced these tests, and
 now show the results of a 3D spherical expansion
  that results in  higher $\gamma$ values than
    the tests discussed in DB02.
  We utilize a cubical  grid of  $100 \times 100 \times 100$ points,
with each side having length unity.\footnote{
  DB02 quote a CPU run time of $\sim2$ minutes per Runge-Kutta
integration for a 3D problem with $100^3$ nodes,
   utilizing the third order reconstruction for the L and R values
 of the primitives, and 
 using    a 1 GHz  Linux PC; our run time of $\sim70$ s is consistent
with the  1.7 GHz  Linux PC used in this work.}
    Initial conditions are that everywhere $v^j=0$,
$\rho=2$,
and $p=3\times 10^5$ within a radius 0.1 of the  center
of the cube, and $p=1$ outside. 
   This is the 3D analog of the
1D  ``piston'' problem.
  (An identical test with less extreme initial conditions
 is shown and discussed by Hughes, Miller, \& Duncan 2002,
  see their Fig. 3.)
  The  enormous over-pressure launches 
                 a relativistic blast
wave at $t=0$.
Figure 1 shows the evolution of $\rho$, $v_x$, $p$, and $\gamma$
for a slice taken along the $x-$axis.
   The density inside the sphere becomes small for late  times.
 To avoid numerical instabilities in this region
   we utilize  the smoother LLF fluxes for small densities ($\rho < 0.0225$),
and the HLL fluxes elsewhere.
  We show the initial configuration plus 
      10 time slices from a run with 360 total time steps, taking
a Courant number  of $0.05$.
   After the run begins  one can see the development of a strong
spherical 
    outflow from the center of the cube.
     The Lorentz $\gamma$ factor  has increased
to $\sim 20-30$ in the outer parts of the expanding shell by the
end of the run. The run was  halted before the  expansion
reached the edge of the grid.
     For the last
three  times steps,
     a kink develops in
the $v_x$ profile  at the center of the sphere because
of the 
    prolonged spherical outflow from localized point.
  Figure 2 shows the corresponding evolution for 
  pure LLF fluxes. All variables now show a smoother evolution
   and  $\gamma \rightarrow 20$ at large
  radii for late times  $-$ somewhat less than for the HLL fluxes.
  Conservation of  rest-mass
       energy $M \equiv \Sigma_i (\rho_i \gamma_i)$
   and energy $E \equiv  \Sigma_i (e_i \gamma_i^2 + p_i(\gamma_i^2 -1))$
 (where the specific energy $e_i = \rho_i + p_i/(\Gamma-1)$)
 integrated over the computational grid is as follows:
  By the end of the run shown in Fig. 1,
  $M$ and $E$ 
 have  increased by about 1 part in $10^4$;
  by the end of the run shown in Fig. 2,
   $M$ has increased by 
 $\sim$3 parts in $10^5$, and $E$ is constant
  (to within machine accuracy).

 \section {  Results  }

We now present results for the case of interest,
namely a relativistic blob of material that expands
   roughly  axisymmetrically (orthogonal  to the direction of motion)
 as it decelerates from ultrarelativistic to relativistic speeds.
 We set up our initial problem as follows:
  our blob is made  to propagate along the positive $x-$axis
   through a 3D rectangular grid
  of  $500 \times100 \times100$ cells.
 The  box measures 5  units along the $x$ axis, and 1
  unit
each along the $y$ and $z$ axes.
    Therefore the grid spacing along each axis is 0.01 units.
  We explore a variety of configurations    
   for our initial state.
 We performed runs in which the blob initially 
is either a small cone, sphere, or plate,
  with symmetry about the propagation vector.
  The dimensions of the initial blob are optimized,
   according to its shape, so that $\sim70$ grid points
%
   out of the $5\times 10^6$ in the 
  computational domain initially comprise the blob.
An initial spread is imparted
 to each fluid element by setting the $y$ and $z$ components
of velocity such that $\theta = (v_y^2+v_z^2)^{1/2}/v_x = 0.035$.
The motivation for taking $\theta=0.035$
                      is the study of Frail et al. 
(2001) that finds  a peak
  in the   frequency histogram
             distribution of inferred spreading angles $\theta$
  from observed GRBs
        of $4^{\circ}$ or 0.07 radian.
 We divide this value by two to obtain the
half-angle spread  $0.035$ radian. The blob is given 
 a Lorentz factor
$\gamma=25$  so that the relativistic
beaming angle $\gamma^{-1}$
   lies within the physical spreading angle $\theta$ of the jet.
%
%
%
  We set  $p_{\rm blob}/\rho_{\rm blob} = 10^{-4}$ (inside the blob),
                  $p_{\rm CSM} /\rho_{\rm CSM} = 10^{-6}$ (outside the blob), 
        and    $\rho_{\rm blob}/\rho_{\rm CSM} = 10^2$.
%
      We also perform one trial  with $\theta=0$, i.e., $v_y=v_z=0$ at all points.
   This is the  ``null hypothesis'' run. 
     The effective density contrast
perceived in the lab (i.e., CSM) frame
  is  $ (\rho_{\rm blob}/\rho_{\rm CSM}) \gamma_{\rm blob}$.
 Soon after the computation starts, the initial profile
   relaxes to one in which  $\rho_{\rm blob}/\rho_{\rm CSM} \simeq 10$,
  therefore the effective density contrast
   for the early run  is $\sim 200$. 
This is less
than expected astrophysically, but required  in our computations in
order to see significant deceleration of the blob  by the time
it reaches the end of the grid $x=5$.
   In other words, we must telescope the evolution
      into  the finite dimensions of our grid.
 From conservation of momentum, the condition for significant
deceleration is that the total swept-up mass-energy roughly equals
that in the initial blob.
%

 Experimentation using HLL fluxes shows
that  strong internal shocks almost immediately
     create large-amplitude  sawtooth $\rho$ variations 
 within the blob, leading to noisy results. Therefore
we utilize the smoother 
              LLF fluxes in this work.
 We  follow the evolution of the blob in terms of
both its motion in $x$ and its spreading in $y$ and $z$.
  Taking a Courant number of $0.25$ necessitates $\sim2000$ time steps
for the blob to reach the end of the grid, traveling at $v_x \simeq 1$.
     This is a simple consequence of the
  box length and  grid spacing $\Delta x = 0.01$,
  from  which it follows that the time
     to traverse the box is $2000 \times 0.25  \times 0.01 =5$  units
  traveling at $v\simeq1$.
 A  density contrast of $\sim10^2$ is sufficient to see the desired
deceleration from $\gamma^{-1} < \theta$ to  $\gamma^{-1} > \theta$
 during a run.

  In  order to avoid undue complexity in these experiments,
    that currently are purely hydrodynamic and do not yet contain
   proper
   prescriptions for emission from bremsstrahlung and synchrotron processes,
     we calculate a simple measure of the emissivity by
   taking the local emission to scale as $p$, which would
  be expected roughly  for optically thin synchrotron emission
  characteristic of frequencies significantly above the 
   self-absorption frequency.
            We avoid the issues
  of synchrotron self-absorption
  (Granot, Piran, \& Sari 1999b, 2000;
   Panaitescu \& Kumar 2000;
     Sari \& Esin 2001),  and 
   of whether the evolution is
 adiabatic or radiative (Panaitescu \& M\'esz\'aros 1998b;
                            Sari, Piran, \& Narayan 1998)  $-$
  our calculations are adiabatic.
         We calculate effective  
  ``light curves'' for observers 
at various viewing angles between $0^{\circ}$ and $15^{\circ}$ 
  from the center of the jet.
   The  amplification of the photon energy flux 
is dictated by three factors:
(1) The rate of photon emission is increased by a factor of 
$\gamma$ as one goes from the blob frame to the lab frame.
(2) The individual photons are Doppler-shifted
  so that their
lab frame energy is increased by a factor of 
$\gamma (1+\beta\cos\phi)$,
where $\phi$ is the angle between the line of sight and blob's 
direction of motion.
(3) The photons are focused in the direction of motion, so
that an angle of emission (relative to the direction of motion) in the
blob's frame, $\phi^{'}$,
  is related to the lab frame
 angle $\phi$ by 
$\cos\phi^{'} = (\cos\phi-\beta)/(1-\beta\cos\phi)$.
The differential solid angle $d(\cos\phi^{'})=
d(\cos\phi) [(1-\beta^2)/(1-\beta\cos\phi)^2] $.
 Thus, 
     the net amplification of the photon energy flux
  for  each fluid element is 
  $(1+\beta\cos\phi)/(1-\beta\cos\phi)^2$,
%
%
  Similar expressions have been derived
   previously
   (e.g., Granot, Piran, \& Sari 1999a, eqns.  [3], [4];
                   Wang, \& Loeb 2001,  eqns. [13], [31]).
 The formalisms of Granot et al (1999a) and Wang \& Loeb (2001)
also contain  a provision for time delays
  based on the relative positions and velocities
    of different fluid elements.
  The ``time'' of interest in summing the relative
    emission contributions is
  the retarded  time $t -r\cos\phi/c$.
  We compute light curves 
  by keeping running totals of the contributions 
  to the flux from 100
 slices through the computational domain
     (straddling the position of local disturbance
  in the fluid caused by the advancing blob)
    that
  are
 orthogonal to the blob velocity vector
  (i.e., cuts in the $y-z$ plane) and  advance
    in time toward the observer at $c$. 
  Over  the course of a run we build up a light curve
  based on constant retarded times with respect to the observer.
 This approach is essential insofar as the photons emitted by
a given fluid element outpace 
  the blob itself only slightly.


Figure 3 shows the evolution
  of physical quantities for the run which begins with a small cone  
               along a cut through
the center of the computational box,
  along the direction of propagation of the blob.
The initial conditions quickly disappear and
 give way to the deceleration phase noted by Kobayashi et al. (1999).
%
  During the entire evolution there is a local maximum in $p$ in advance
of the propagating blob  that accompanies the leading pile-up
in density. The region of  highest $\gamma$ lies within the
density minimum  that lies just behind this shocked region.
 Although the initial pressure is small, after the bow shock
is fully developed we have $p_{\rm shock}/\rho_{\rm shock} \simeq 0.3$
   throughout the subsequent evolution.
       The evolution proceeds
  in a roughly self-similar manner and the
  forward/reverse shock system is stable,
  in accord with analytic estimates
   (Wang, Loeb, \& Waxman 2002).

Figures 4 and 5 show 
    a time series of contour plots in $\gamma$ and $\rho$
 that follow the blob propagation and radial spreading.
    The full 500 grid
points along the $x$ direction are shown, and each frame represents
 240 time steps (i.e., $\Delta t = 0.6$).
 The leading contours show the density enhancement associated
with the shock. The $\rho$ contours indicate the values 
 0.2, 0.3, 0.4, and 0.5, 
    while the (trailing)
$\gamma$ contours,
indicate values of 1.25, 2.5, 5, 10, 15, and 20.
    The contours are formed by taking a cut at $z=0$ through
the $x-y$ plane. The initial velocity vector of the blob
points toward the top of the plot, with $\gamma=25$.
 One
sees a strong bow shock associated with the
maximum in pressure at the point where the blob encounters the
  CSM.  The lack of a bow shock in the $\gamma$ contours
 suggests that  little or no material in advance of the blob
is accelerated to a significant bulk Lorentz factor.
   As the evolution progresses the deceleration
       causes the higher $\gamma$ contours
to disappear,
              and those that remain become increasingly distorted.
   The bilateral symmetry of the contours evident in Figures 4 and 5
                  attests to the power of the third order differencing scheme
  given in DB02, insofar as the 3D model
  has no enforced symmetry.

Figure 6 shows the evolution of the total  rest-mass energy $M$
 (dotted) and 
total energy $E$ (dashed)
     within the grid, over the course of the conical run.
   The values of $M$ and $E$ have been normalized to their
initial values.
 The curves that  extend significantly below an abscissa
value of unity indicate the values summed over the grid.
   At the six faces of the computational box we continually 
reset the  values of all variables to their initial values, 
and keep track of the differences between those values 
and the initial ones. As the run progresses and
more high velocity gas reaches the edge of the computational
domain
 and is extracted, there is a net extraction of
  positive   $M$ and $E$ from the grid. In addition, 
   there is a low density wake that  trails the blob, 
and therefore at the contact points between the wake and the
sides of the grid  there is a net extraction of negative
  $M$ and  $E$. From a practical standpoint,
   the density inside the wake 
  is so small that 
   the net extraction of negative
   $M$ and  $E$
   does not significantly affect  the bookkeeping.
            The curves
     in Figure 6 that
  lie  close to an abscissa value of
unity indicate the summed values for $M$ and $E$ corrected
 for both net extracted positive and negative rest-mass energies.
  The second panel of Figure 6 gives an expanded view of the first
panel 
 around the abscissa value of unity.
    After doing the bookkeeping on the $M$ and $E$ in the
  extracted gas, conservation of total rest-mass energy
and  total energy are good to within $\sim1-2$ parts in $10^3$.

Figure 7 shows  values of total rest-mass energy within 
selected high $\gamma$ cuts through the computational domain
 for the initially conical run,
  and the evolution of the 
  $x-$component of momentum $<\gamma v_x>$ for five different runs $-$
 the initial blob as a plate  (``P1'' and ``P2''),
   sphere (``S''), cone (``C''), and cone with zero spreading (``$v_T(0)=0$'').
  The weighting function used in evaluating  $<\gamma v_x>$
  is $W(\gamma) = (\gamma^2-1) \rho\gamma$.
   The run P2 uses fewer grid points ($500\times 50^2$)
  than the other runs, thus material
    shunted aside by  the bow shock  reaches the edges of the computational
   domain earlier. In spite of this, the evolution of P1 and P2 is quite similar,
   i.e.,  there is  no
  dynamically significant back reaction on the material inside
  the computational domain induced by the departing fluid.
 As we detail in the Discussion section, the drag force on the 
     blob increases as the effective area presented by the blob,
 which for a roughly constant lateral spreading goes as $x^2$.
 Therefore the integral of the relativistic  impulse
 equation $d (M_{\rm blob} \gamma_{\rm blob} v_x)/dt = F_{\rm drag}$
        provides a solution of the form 
   $<\gamma v_x> \simeq a - b(x + c)^3$,
 where our fitting for the conical run
    ($a=21$, $b=0.15$, and $c=0.45$)
 is shown by the curve labeled ``$f(x)$''.

   The curves in Figure 7 showing the variation
  of the rest-mass energy contained within
 different $\gamma$ cuts follow the values
    $\gamma_{\rm cut} =$ 1.001, 1.01, 
  1.1, 1.5, 2, 4, 6, and 8.
  The higher $\gamma$ cuts $\gamma_{\rm cut} \ga 2$ 
   reveal
    that a  negligible fraction of
   CSM matter gets accelerated to significant values.
 This is because the curved bow shock shunts material laterally
     in front of the advancing blob, rather than accelerating
  it up to a significant fraction of the blob's bulk Lorentz factor $\sim10-20$.
  The rest-mass energy  curves are  relatively constant
  in time
up to $t\simeq 2$. 
 At roughly  this point
   the total accumulated CSM gas  becomes comparable to
that in  the initial blob.
  If the blob
    did not spread laterally,
  this time would occur 
 when the
 blob had plowed through
  a density $\sim \gamma_{\rm initial} (\rho_{\rm blob}/\rho_{\rm CSM})$,
  where  $\gamma_{\rm initial} = 20$ 
 and $\rho_{\rm blob}/\rho_{\rm CSM}\simeq10$.
  This means that the blob would
     have shed of order its initial momentum
      by the time it had
traveled $\sim 200$ times its initial length $\sim\delta x = 0.1$,
  or    $\sim20 $ units.
  In practice, the  initially imposed
   tangential velocity $v_T$
    leads to the 
   lateral expansion that
     effectively increases the
cross section for interaction of the blob.
      The negligible decrease in $<\gamma v_x>$ 
      for 
       the $v_T(0)=0$ run demonstrates the importance of this effect.
%
%

   Figure 8 shows the lateral expansion of the
blob  for the conical run.
   The blob edge   
     is computed in each time step by
       first finding   the
local maximum in either $\rho$, $\gamma$, or
     $\rho\gamma$  along the $x-$axis,
   and  then
stepping laterally to the position at which the 
  background (CSM) value
  of the relevant quantity  has been 
         increased by 10\% due to 
 expansion of the blob.
   The transient lateral
  expansion  $v_{\rm edge} \simeq 0.3-0.4c$
          lasting until  $x\simeq 1$
       is unphysical insofar as it
   occurs during
the period of adjustment to the 
initial density and velocity profiles.
   The value of the  later
     spreading rate  $v_{\rm edge} \simeq 0.1c$
  is basically dictated by the initial $v_T$ value
    given to the gas.
%
%
%
%

Figure 9
       shows the evolution of $<\gamma v_x>$ and $\theta$ as the
blob propagates. The solid curve in the top panel indicates
the same weighted value of $<\gamma v_x>$ as shown in Fig. 7.
     After the
transient physical conditions associated with the
    initial state have vanished,
  one sees a period of deceleration
   associated with the increasing
   drag force.
%
%
The dashed curves  indicate the reciprocal of the full width
spreading angle determined by summing and averaging
the local values of $2(v_y^2 + v_z^2)^{1/2}/v_x$.
   Due to the complicated  pattern of spreading and the mismatch
between contours of constant $\rho$ and constant $\gamma$, 
 the specific value for $<\theta>$ depends on where one cuts
off the outer edge of the  blob.
  The three  dashed curves use limiting lower values of 4, 5 and 6,
respectively, for the $\gamma$ value, in determining the volume
averages that enter into $<\theta>$.
  The spherical/jetlike transition
occurs  when $\gamma_{\rm max}$
  drops below the $<\theta>$ value computed using
the lower limit $\gamma_{\rm cut}$ = 6.
 The bottom panel shows the number of cells used
in  computing  the $<\theta>$  values.
  At  late times the disappearance of higher velocity
matter makes problematic a calculation of $<\theta>$
based on cuts in $\gamma$.

Figure 10 shows the evolution of our canonical ``luminosity'' measure,
namely $\Sigma_i p_i (1+\beta_i\cos\phi)/(1-\beta_i\cos\phi)^2$,
%
for various viewing  angles $\phi$.
    As noted previously,
  the light curves are calculated by summing the emission
   from all points with the computational domain
   that lie, in a given time step, on 100 $y-z$
  planes which straddle the position of the blob
  and advance toward the observer at $c$.
%
%
 One sees a slight
    break in the ``light curve''
   for the $\phi=0^{\circ}$
  observer 
  at 
  $\Delta\log t(x) \approx -0.3 $ 
   due to the
   transition from spherical to jetlike expansion.
 Although our model is too simple to allow any detailed
   quantitative comparison with observations
   at this point, it is worth noting that many 
    of the observed afterglow transitions are also 
   smooth
   (e.g., Fig. 2 of Stanek et al. 1999;
       Fig. 1 of Harrison et al. 1999),
    rather than the abrupt breaks evident in some
   of the semi-analytical models.
  The transition is not so visible for off-axis
   viewing angles. 
%
  The general pattern in which increasingly off-axis
  observers see significantly lower
  emission until well past  maximum  light curve 
can be seen in the light curves
shown in Fig. 4b of 
 Panaitescu \& M\'esz\'aros (1999).
  Afterglows coming from
those GRBs  for which the
  observer does not lie within the 
   initial cone of GRB emission
   are referred to ``orphan''afterglows.
 Future surveys of orphan afterglows
 might provide constraints on the
  beaming that are different from those
of Frail et al. (2001, e.g., Totani \& Panaitescu 2002;
                Levinson, Ofek, Waxman, \& Gal-Yam 2002).

\section {Discussion}

   Utilizing 3D relativistic hydrodynamical 
  calculations,
 we have examined the  evolution of an expanding
relativistic blob of gas intended to be representative
of a  jet  associated with ejecta 
 from an extremely energetic event such as a hypernova,
 that produces a gamma-ray burst (Aloy et al. 2000;
  Tan, Matzner, \& McKee 2001;
  MacFadyen, Woosley, \& Heger 2001,
  Zhang,     Woosley, \& Heger 2003,
  Zhang,     Woosley, \& MacFadyen 2003).
     Since these are the first such calculations
   applied to the blob during the time in which 
the afterglow radiation 
        is produced,
  we have purposely  kept  them simple
  in an effort to concentrate on 
  the most fundamental aspects of the physics.
  We restrict our attention to 
     the transition
from   spherical to jetlike expansion that
occurs during the time that the Lorentz factor
 becomes less than 
     the reciprocal of the  jet spreading angle.

We have not yet attached  specific numbers
 to our results. From the SRHD equations,
   one
sees that the relevant quantities are the 
ratios of pressure to density, and of distance to time.
   If we specify either one of these two sets of
numbers, the other one is also determined.
  The column giving the observed afterglow 
break time $t_j$ in Table 1 of Frail et al. (2001)
 indicates $t_j \simeq 2$ d as being representative.
 For an observer directly on the velocity 
   vector of the blob, 
  the time $T$ between the GRB and afterglow

$$ T = \int \delta t = \int_{t_{\rm GRB}}^{t_{\rm afterglow}}
                 \left({\delta x\over v(x)} - {\delta x \over c}\right)
 =
   {1\over c} \int \delta x
      \left[ { 1 \over \sqrt {1- \gamma^{-2}} } - 1 \right] \simeq {1\over 2c} 
   \int \delta x \ \gamma(x)^{-2}, $$

\noindent
  where the dominant contribution to the integral comes from later times.
   Thus the light travel time of 1 day  is multiplied by $\sim 2 \gamma_{\rm afterglow}^2
  \simeq 2\times 10^2$, assuming the spherical-to-jetlike
                       transition giving the break in the afterglow 
 light curve happens at $\gamma \simeq 10$.
  For the conical run,
  the break in the light curve
   occurs at
%
%
   at $x\simeq 2.5$.
 If we designate this point as corresponding to
 a time 2 d, then  $x=2.5$ translates to $c t_j \simeq 5.2\times 10^{15} (200)$ cm $\simeq 10^{18}$ cm.
 Since we have 100 grid points per unit length, this 
means  each grid point spacing corresponds to $\sim 1 \times 10^{16}$ cm.

Zhang, Woosley, \& MacFadyen (2003)
     utilize a 2D relativistic hydrodynamics code
   to follow the evolution of the blob through the
  envelope of the progenitor star
   and past the point of ``break-out'' through the stellar
surface.
   They find that a blob launched at $r = 2000$ km 
 with an initial spreading angle of $\theta_{\rm initial} = 20^{\circ}$
  and a bulk Lorentz factor of 50
   experiences strong shock heating and also lateral 
confinement,  so that when it emerges
   from the progenitor it has a large internal
  energy $p/\rho\simeq10$, a bulk Lorentz factor $\sim10$,
     and is more confined than
     initially ($\theta_{\rm final} \sim5^{\circ}$).
    The subsequent expansion due to the large $p/\rho$
   value leads to an effective $\gamma\simeq2\gamma_{\rm bulk}\gamma_{\rm thermal}
  \simeq 200$
    for a distant observer.
In the standard model for GRBs, the
gamma radiation is produced by strong internal
shocks in the expanding fireball  
  at the point where it becomes optically thin
to its own radiation, at $\sim 6 \times 10^6 \ (E_{51}/n_0)^{1/3}$ s,
where $E_{51}$ is the total energy carried by the jet, corrected
for beaming, in units of $10^{51}$ ergs, and $n_0$ is the CSM density
 in units of $1$ cm$^{-3}$. During the time of gamma ray emission,
the bulk Lorentz factor of the 
  ejecta  $\gamma\simeq 100-300$.
   Zhang, Woosley, \& Heger (2003) compute the evolution 
  after  break-out covering the time 
  of the prompt GRB emission. They estimate it should occur
at about $3\times 10^{14}$ to 
        $3\times 10^{15}$ cm and have a duration
   $\sim r/(2\gamma^2c) \simeq 10-10^2$ s.
     In this work 
   we focus on the time after this event, 
    i.e., subsequent to the evolution considered by Zhang et al. (2003ab),
when the ``afterglow'' is produced. This emission is
thought to arise primarily in the bow shock where the 
strong heating leads to $p_{\rm shock} \simeq \rho_{\rm shock} $
      (Wang \& Loeb 2001).
 In essence, this energy production arises from ``external shocks'',
as opposed to the ``internal shocks''
  that generate the gamma radiation 
    (Rees \&  M\'esz\'aros  1994;  Panaitescu, \& M\'esz\'aros 1998a;
   Spada, Panaitescu, \& M\'esz\'aros 2000,
    Kobayashi \& Sari 2001).

 After the blob is ejected from the envelope of the 
star that was the hypernova precursor
   and continues to  propagate,
   the medium through which it travels
  should be dominated by the density profile
  left from the precursor's stellar wind 
     (Chevalier \& Li 1999, 2000, Li \& Chevalier 2001).
 If the density varies as $r^{-2}$ away from the star,
 the mass loss of the wind ${\dot M}_{\rm wind} \simeq 10^{-5}\msunyr$,
 and the wind velocity $v_{\rm wind} \simeq 10^3$ km s$^{-1}$,
  the circumstellar density at the point where the afterglow
             radiation is emitted 
     $\sim 10^{18}$ cm will be 
   $n_{\rm CSM} \sim  {\dot M}_{\rm wind}/(4\pi r^2 v_{\rm wind} m_p)
         \simeq 0.3 $
         cm$^{-3}$.
In the calculation of Zhang et al (2003), the density
inside the jet  is $\sim 10^{17}$   cm$^{-3}$  at $\sim 10^{12}$ cm,
and $p\simeq\rho$.
  If the spreading angle of the jet 
 remains  roughly constant
     from  $\sim 10^{12}$ cm to  $\sim 10^{18}$ cm
 (Lithwick \& Sari 2001), 
      then
      the density inside the jet at $\sim 10^{18}$ cm
should be lower by $\sim (r_2/r_1)^2 \simeq 10^{12}$,
or about $\sim 10^5$  cm$^{-3}$.
 At this point $p$ inside the jet should be
negligible compared to $\rho$.
  The density contrast between the blob and CSM is greater
than what we have assumed in the results shown previously.
    A separate run taking an initial density contrast
 of  $10^6$ shows the same  basic effects as the previous runs, however,
  namely a lateral expansion rate of $\sim 0.1c$ and the
  non-accretion of CSM gas.

  Our computations lie within  the ``deceleration
  phase''  discussed 
 by Kobayashi et al. (1999,
 see also Kobayashi \& Sari 2000, 2001).
We find a  change in the form of the luminosity
decrease corresponding to the
transition between spherical and jetlike 
    expansion.
   The determination of 
 the average spreading angle $<\theta>$ is nontrivial
because it depends on how one does the averaging,
and how much of the diffuse, sideways-expanding 
jet material is included in the computation.
  In Figure 9  we presented cuts
 for  gas  possessing  $\gamma > \gamma_{\rm cut} = $ 4, 5, and 6
as representative of material in the flow  that
partakes most strongly in producing the observed
radiation. 
   We do not see a dramatic increase in  $<\theta>$
during the deceleration phase; 
    the  $<\theta>$ value basically reflects the ballistic
     motion of material following its initial $v_T$ value.
%
  Also, because the  deceleration in our problem
is forced not by the accumulation of gas from the CSM 
but rather the drag force of the CSM on the blob,
       we do not
see an exponential decrease in $\gamma$ with distance
during the deceleration phase, but rather a
decrease well-approximated by a cubic dependence
   that is expected for a roughly constant lateral spreading rate.

   Many of the previous studies, utilizing the formalism of BM76
in which matter accreted  from the CSM 
  is added onto the GRB shell and facilitates the deceleration,
   have discussed the sideways expansion of the
   jet that supposedly occurs near the time that most of
the initial kinetic energy of the blob has been shed.
(A qualitative depiction of this effect is shown  in Fig. 1
 of Piran 2002.)
   Our results do not appear to support this viewpoint.
       We find that
  the CSM gas hitting the bow shock does not add significantly
     to the 
mass of the jet, but rather is swept back into its  wake
       as the blob passes a given location. 
  In this sense,
  the relevant factor in determining
the deceleration
    is simply the projected  CSM surface
density relative to that in the blob.
   As a result,
      there is no runaway phase of rapid lateral
expansion $-$ a physical process 
      that has been ``built in'' to
   some previous theories. 
   We do not see an abrupt jump
  in the lateral spreading coincident
     with the   deceleration   phase;
   the lateral spreading rate is mandated basically by the
   initial $v_T$ value imparted to the gas.
   The pressure-driven lateral expansion is negligible
    until such time as $v_{\rm spread} < c_s/\gamma$,
   which for $c_s \simeq v_{\rm spread}$
            is so late as to be uninteresting.
     In the aforementioned works
    there is a close connection between
           the transition  $v_{\rm spread} \sim c_s/\gamma$
  and the change from spherical to jetlike emission, whereas in 
   our study there is not.
     The  break   in the light curve
        produced by emission from the bow shock
  is due solely to 
      the lessening of 
relativistic beaming brought about by a slow, uniformly 
  increasing drag from the CSM,
  and the fact that once $\gamma^{-1}$ becomes less than $\theta$,
   the observer perceives the finite width of the jet.

    Rhoads (1999)
   estimates that the lateral spreading speed of the
jet should be $v_{\rm edge}\simeq c/\sqrt{3}$.
   Sari, Piran, \& Halpern (1999)  
  argue for a faster spreading rate $v_{\rm edge}\simeq c$.
   Rhoads (1999)
divides the 
 dynamical evolution  of the blob into two regimes, 
characterized by a power law decay (in time) of the bulk Lorentz
factor, followed by an exponential decay.
During the latter stage, the swept-up mass increases
exponentially in time.
  Panaitescu \& M\'esz\'aros (1999)
  calculate light curves for observers
at varying angles from the  jet axis, and calculate
separately the effects of including and excluding the
lateral jet expansion.
  They  find that the maxima in the light curves
occur substantially later in runs which do not take into
account the jet broadening (see their Fig. 4).
  In the analytical models of both  Rhoads (1999) 
and Panaitescu \& M\'esz\'aros (1999),
the physics of mass accumulation from the  
  CSM is an integral component of the formalism;
 all mass within the solid angle of the
expanding shell is assumed to accrete. 
 %
      Workers have applied the results of 
  Rhoads (1999) and  Sari, Piran, \& Halpern (1999)
 to the afterglow evolution, however  the results of Zhang et al. (2003) 
  cast doubt on the validity of this exercise,   because at the time
corresponding to the afterglow emission one anticipates that  
                 $p << \rho$ and the lateral spreading rate
       would not be governed by the internal sound speed  but rather
       the ballistic motions of the ejecta comprising the blob
     as they leave the vicinity of the progenitor star.
   The blob has a large  internal thermal energy
   $p/\rho \simeq 10$ as it emerges from the progenitor star,
    and even by the time the expanding ejecta have become 
 optically thin to their own  emission from internal shocks
   (producing
    the GRB), one still expects $p/\rho\simeq 1$.
  By the (much later) time of the afterglow emission,
          however, the blob
   would have cooled to the point that  $p/\rho <<1$, which
   provides the impetus for our initial condition
                $p_{\rm blob}/\rho_{\rm blob} =10^{-4}$. 
  Following the arguments of Rhoads (1999) and Sari, Piran, \& Halpern (1999),
   if the lateral spreading
  rate were mandated by the internal sound speed, then in our
  calculations it should be $\sim \sqrt{p/\rho}  = 0.01c$, 
   whereas we find it to be $\sim10$ times larger.
   The spreading rate  basically  reflects our initial $v_T$ value.
%
     Thus  the physics of the lateral expansion is different
   than in Rhoads (1999)  and  Sari, Piran, \& Halpern (1999).
    How might this result  be influenced by systematic
   effects present in our model? 
    One obvious potential shortcoming is the absence of
   cooling.
      In this work we have assumed an adiabatic gas, whereas
        in reality one might envision
        the presence of cooling within the
        shock.
    This might then reduce the ability of the shock to 
   deflect the gas as efficiently as it now does,
   which in our model prevents
         the acceleration of CSM material to Lorentz factors
    approaching that of the blob.

   We 
adopt a relatively
small density contrast
         between material inside the blob
and the exterior CSM
      in order to see
significant evolution of the blob during the course
  of the simulation within our 
 Eulerian grid.
      We have run additional models using 
   $\rho_{\rm blob}/\rho_{\rm CSM} = 10^6 - 10^8$,
     and although these could not  be run
  for sufficient time to see the deceleration phase
begin, we find in these runs also  a
lateral spreading  rate $\sim0.1c$, for the same initial $v_T$ values,
     and negligible accretion  of gas  from the CSM.

    We may understand the
  deceleration
of the blob with
  a simple analytical model. 
      If we
      assume that the only significant
     contribution to $\gamma_{\rm blob}$ is through
forward motion, that the
drag force varies as $v^2$, 
  and that $p_{\rm blob}<<\rho_{\rm blob}$,
   then $ \partial_t (M_{\rm blob}\gamma_{\rm blob} v_x) = F_{\rm drag}$ implies

$$\partial_t (M_{\rm blob} U)= - K v_x^2  \sigma \rho_{\rm CSM}/c,$$

\noindent
where $U\equiv \gamma_{\rm blob} v_x/c$, 
         $\sigma$ is the cross sectional area
         for interaction $\pi (v_{\rm spread} t + r_0)^2$,
     $v_{\rm spread}$   is the (constant)
  lateral spreading speed ($\sim0.1c$ in our calculations),
 and   $K$ is a  dimensionless number
   (typically of order 0.2 in laboratory
  applications).
     Taking  
   $\partial_t = v_x \partial_x$ and $c=1$ gives

$$\partial_x (U)= - K v_x  \sigma \rho_{\rm CSM}/M_{\rm blob},$$

\noindent where 
  the blob mass $M_{\rm blob} = \gamma_{\rm blob} \rho_{\rm blob} V_{\rm blob}$.
  The blob volume is given by the number of cells comprising 
  the blob initially ($\sim70$) times the volume of an element
   $\Delta x \Delta y \Delta z = (0.01)^3 = 10^{-6}$.
Therefore $M_{\rm blob} = (25)(10^2)(70)(10^{-6}) = 0.17$.
Integrating gives

  $$U(t)   =  U_0  - a (x+b)^3 v_x,$$
  
  \noindent
  where $U_0$ is the initial value 
     of $U$ and $a = (K/3) \rho_{\rm CSM} \sigma /M_{\rm blob}$.
 From the   fitting to $<U>$ presented 
                 in Figure 7,
   we infer that $K \simeq 2.4$.
       The specific numerical value for $K$ is probably influenced
by our numerical resolution.
   A comparison of the curves labelled ``C'' and ``$f(x)$''
  in Fig. 7 shows that the functional decrease in  $<U>$ with $x$
   is reasonably described by a cubic, as expected
   if the cross sectional area increases  quadratically with $x$,
          or equivalently
   $t$.
    Note that we assumed $v_x=1$ in this exercise, which is a good approximation
   for the evolution of interest.

%
%
%
%
%

 
%
%
%
%

\section {Conclusion }

The calculations we present
  are the first
               3D relativistic hydrodynamical calculations
of GRB jet evolution  pertinent to the afterglow phase
                 that do not enforce any special
symmetry (e.g., spherical or axial).
%
%
%
%
      We find that
    (i) 
      the CSM gas does not accrete onto the advancing blob,
        but rather is shunted aside by the bow shock,
  (ii) 
      the decay light curve
      steepens
      roughly when one first ``sees'' the edge of the jet
 $\gamma^{-1} \approx \theta$,  with this effect being
   strongest for ``face-on'' observers
  (confirming previous studies), and
%
%
(iii) the rate of decrease of the $x-$component
  of momentum $<\gamma v_x>$ is well-characterized by a simple
  model in which the cross sectional area of the blob increases
     quadratically
             with laboratory time (or distance).
  The primary impetus for the
   built-in
    assumption of  accretion of matter
     in previous 
      studies was the influential work of
        BM76 in which
      spherical relativistic  expansion was considered.
    Accretion of gas
   onto the relativistically expanding shell
      is obviously justified
  for spherical expansion, but
 subsequent GRB workers applied the results
  to  the case of the GRB jet, in which a thin wedge of
material propagates through a low density medium.
     In such a situation the natural tendency of material
in front of the jet is to be pushed aside and to form
a ``channel flow'' around the jet,
                            rather than to accrete.
 A separate issue 
  is that workers
   used the results of Rhoads (1999) and Sari et al. (1999)
   that give  lateral expansion rates of order $c$
     to explain the phase of afterglow evolution,
   while it now turns out from the work of Zhang et al. (2003ab)
   that the much smaller internal energy in the
   afterglow-producing  blob would be
expected to give a much lower spreading rate,
   were that the only relevant spreading
                mechanism.
      The evolution of the blob prior
   to the afterglow epoch, however, in particular
  its emergence from the progenitor envelope,
   did encompass a period of much greater $p/\rho$
   during which a large $v_{\rm spread}$  was
  imparted to the gas.
  In summary, there is nothing in our results
       to suggest that BM76, Rhoads (1999), and Sari et al. (1999)
   are not internally consistent, rather it appears
   that the subsequent application
   of their results to afterglow evolution
 may have been inappropriate.

       Recent semi-analytical work has centered
       on 
     ``structured''
           relativistic jet modeling, 
   wherein one replaces the ``top-hat'' assumption
  of uniform jet properties, e.g.,
  Lorentz factor and $p/\rho$ constant
  within the initial cone angle,
   with more physically motivated forms in which
   the physical parameters vary 
   with angle away from the jet symmetry axis 
  $\phi$ (e.g., Kumar \& Granot 2003, Granot \& Kumar 2003).
 Structured jets represent
   a refinement in the
  semi-analytical and analytical work
   in the sense that unphysical artifacts
  of the top-hat models are avoided.
  In this work
    we allow naturally occurring gradients within
  the flow determine the evolution. In comparing the runs
  with different starting conditions,  for instance,
   we see that only the gross initial characteristics
  (bulk Lorentz factor and spreading rate)
      are of
  relevance. One could impose a structured initial  state, in terms
  of having an  angle-dependent relation between, for instance
  Lorentz factor and $\phi$, and/or $p/\rho$ and $\phi$,
           but one suspects that the initial intricacies
   would be washed out, just as the initial
  blob shape is.   One might in fact imagine
   inverting the problem
    and  using our results
   to determine physically  motivated functional forms for
   $\gamma(\phi)$ and $p(\phi)/\rho(\phi)$
   as input to a structured jet formalism, but
                  in our models
      the detailed evolution
   of the physical parameters and the interaction of the blob with the
bow shock reveal gross changes along
 the jet axis.  Therefore the motivation 
                              for trying to characterize
    the jet properties as simple functions of $\phi$, averaged
   along the axis of the jet, seems questionable.
 In addition, one would still be missing important physical effects, 
   such as the non-accretion of the forward CSM material.

Two obvious
 refinements, currently being carried out, 
                 are to
(i) treat the problem on a Lagrangian grid  in which
  the mesh points follow  the blob and are
adaptively inserted in regions with strong gradients,
     so as to be able to  explore regimes
  in which the density contrast between  the blob and
  CSM is much larger, and
(ii) include provisions for realistic  bremsstrahlung
  and synchrotron physics, in order to produce
  light curves that can be compared directly with observations
 so as to test different aspects of the theory  and
thereby constrain the allowed parameter space.
 

Our sincerest thanks go to Luca Del Zanna who provided
key insights into the workings of DB02.
   We also thank
   John Baker,
    David Band, Tom Cline,
     Chris Fragile, Chris Fryer,
   Markos Georganopoulos, Peter Goldreich,
     Demos Kazanas, Pawan Kumar, Zhi-Yun Li, 
  Andrew MacFadyen,  Peter M\'esz\'aros,
   Ewald Mueller,
     Jay Norris, Alin Panaitescu,  Sterl Phinney,
    Steve Reynolds, James Rhoads,
    Steve Ruden, Sabrina Savage,
   Craig Wheeler,
   and  Weiqun Zhang for helpful comments.

\vfil\eject
\centerline{ FIGURE CAPTIONS }

Figure 1.
 The evolution of a spherical relativistic expansion
due to a large over-pressure  inside a sphere of radius 0.1,
 adopting HLL fluxes.
  The evolution encompasses 360 time steps, taking a
Courant number of 0.05. 
  Shown are  the initial conditions plus ten equally spaced
time steps taken from a slice along the  $x-$axis
     depicting 
  the evolution of
 (i)
  pressure  $p$ ({\it top panel}),
  (ii)
 density $\rho$ ({\it second panel}),
 (iii)
   $v_x/c$      ({\it third panel}),
 and
  (iv) 
    Lorentz factor $\gamma$ ({\it bottom panel}).
 The small numbers beside each curve indicate
  the evolution.
 One sees an expansion of the central over-pressurized
sphere into the surrounding medium. The strong relativistic
 outflow peaks at $\gamma \simeq 20-30$ near the end of
the evolution. The general trend in which $\gamma \propto r$
within the expansion can be derived from fundamental principles
and is well known from classical solutions.
  At late times the strong decrease in $\rho$ at the center
of the sphere, which is a singularity in this test, 
begins to   cause numerical instability.
  We utilize the LLF fluxes for  $\rho < 0.0225$ and
the HLL fluxes for  $\rho \ge  0.0225$.

Figure 2.
 The evolution of a spherical relativistic expansion
due to a large over-pressure  inside a sphere of radius 0.1,
 adopting pure LLF fluxes.
  Quantities shown are the same as in Fig. 1.
  The evolution is nearly identical to that in Fig. 1,
  except the variations in the physical variables are
smoother, and $\gamma \simeq 20$  at larger radii by the
end of the run $-$ somewhat smaller than in Fig. 1.

Figure 3.
 The evolution of a relativistic blob launched
as a small cone   traveling along the $+x$-axis, initially
confined to between $x=0.1375$ and $x=0.2$
and a maximum radius 0.025.
   The small numbers beside each curve
                  indicate the time step.
  The panels are the same as shown in Fig. 1.
  Snapshots  represent conditions along a slice
   through the center of the rectangular grid,
          taken every 120  time steps ($\Delta t=0.2$).
For this trial and all that follow
       we utilize a Courant number of 0.25 and pure LFF
       fluxes.
       The initial bulb Lorentz factor
       $\gamma$ was set to 25, and the initial
full width flaring angle to 0.07 radian
   as measured by $\sqrt{v_y^2 + v_z^2}/v_x$.





Figure 4.
  The  spreading in $\rho$ in the $x-z$ plane.
  The eight panels depict the initial state, and 
seven subsequent snapshots taken every 240 time steps (i.e., 
                                    $\Delta t = 0.6$).
  The evolution represents a slice through the midplane of
the grid, and the full length and width of the box are shown.
%
%
    Contours indicate values of
 0.2, 0.3, 0.4, and 0.5 (in the same dimensionless
   units as shown in Fig. 3).

Figure 5.
  The  spreading in $\gamma$ in the $x-z$ plane.
The conventions are the same as in the previous figure.
  Contours
indicate values of 1.25, 2.5, 5, 10, 15, and  20.  At later times 
the higher $\gamma$ contours
disappear due to deceleration.
  The absence of a bow shock in this depiction shows that
 material in front of the advancing blob is not accelerated to
$\gamma$ values much above unity.

Figure 6.
   Conservation of total rest-mass energy $M$  and total energy $E$.
  At $t \simeq 1.6$  material starts to 
reach the edges of the grid and flow off, therefore
   one sees the start of a linear decrease in $M$  and $E$.
The curves which closely follow  $y=1$ show
  the values for  $M$ and $E$
after  correcting for material extracted from the grid.
   The bottom panel shows an expanded view of the
region near  $y=1$ in the first panel.

Figure 7.
  The evolution of rest mass energy contained within
various mass cuts, for the conical run (solid lines),
     and the evolution of the weighted
$x-$component of the momentum $<\gamma v_x>$, 
             divided by 5 so as to
be on a common scale with the rest mass energy curves (dashed lines).
  The numbers associated with the solid lines
   indicate the $\gamma$ value used in each rest mass energy cut.
We  show the $<\gamma v_x>$
  evolution 
    for the conical run (C),
      the spherical run (S) 
  and the two plate runs
(P1: $500 \times 100^2$, P2: $500 \times 50^2$).
  For these four runs the initial spread in velocities
is such that $\theta=4^{\circ}$,
whereas for
      the run labelled $v_T(0)=0$ 
  there is no tangential component initially to the 
velocities. Consequently the reduced lateral spreading
rate leads to a slower $<\gamma v_x>$  decrease.
   The 
    function $f(x)$ represents  a fitting to the 
conical run given by $ 21 - 0.15(x+0.45)^3$.
   Although the rest mass energy curves for $\gamma_{\rm cut} = 1.001$
and 1.01 show a sharp increase (up until $t\sim 1.7$ when
material starts to leave the edges of the grid),
   the near constancy of the $\gamma_{\rm cut} = 2$
  curve and the slight decline for higher  $\gamma_{\rm cut} $ 
  curves indicate that there is negligible acceleration of  
  high$-\gamma$ material.
   Also, the functional form of the decrease in $<\gamma v_x>$
  is similar for all runs in which $\theta=4^{\circ}$
  initially. 
%

Figure 8.
   The location of the edge of the jet, determined by
first finding the location along the
direction of  propagation of the maximum in 
   either $\gamma$      ({\it top     curve}),
          $\rho\gamma$  ({\it middle  curve}), or
          $\rho$        ({\it bottom  curve}),
then stepping laterally to the point at which the
background value has increased by 10\% of its original value.
  (The precise value of this constant does not affect the results.)
  The initial maximum radius of the blob is 0.05.
  After $x\sim1.8$ the two lower curves coincide.

Figure 9.
  The evolution of the weighted mean of the $x-$ component
  of momentum
 and a measure of the spreading angle
 $<\theta>$,
     taken to be the ratio of tangential to axial speeds,
   in the observer's frame.
 Shown are
    (i) $<\gamma v_x>$  (solid line) and
 the reciprocal of  $<\theta>$, 
                 measured in radians (dashed lines),
where the lower limiting  $\gamma$ value used in the averaging for
$<\theta>$ is taken to be either 4, 5, or 6 ({\it top panel}).
   (ii) the  $<\theta>$ values  whose reciprocal values are
indicated in the first panel ({\it middle panel}),  and
  (iii) the number of cells entering into the averages for
the three  $\gamma_{\rm cut}$ values in the first two panels ({\it bottom  panel}).
  At late times the deceleration makes the  $<\theta>$
calculations problematic because the number of high $\gamma$
 cells drops.
  This progression in the loss of high $\gamma$ cells
           can be seen at $t\ga 4$
in the bottom panel. 

Figure 10.
 Light curves constructed by summing the
quantity    
 $p_i(1+\beta_i\cos\phi)/(1-\beta_i\cos\phi)^2$
  over the grid.
 We take this global sum as being a measure of
the luminosity seen by an  observer looking down the jet.
  The light curves are built up during the course of the run 
   by summing the emission
from 100 slices moving toward the observer
at $c$ that
 straddle the location of the disturbance
 induced by the blob.
  Thus the light curve consists  of a maximum of 100 points,
where $\Delta t$  is measured with respect to the arrival time
of the first comoving slice.
  The five  values of the angle  $\phi$  between the observer
   and the local velocity vector of a fluid element are
  (going to smaller emission values)
 $0^{\circ}$,  $3^{\circ}$,  $6^{\circ}$,  $10^{\circ}$, and  $15^{\circ}$.
 For the ``face-on'' viewer (inclination $ = 0^{\circ}$),
   there is a mild change in the decay slope, i.e., break, 
  between the regimes of ``spherical'' and ``jetlike'' expansion.

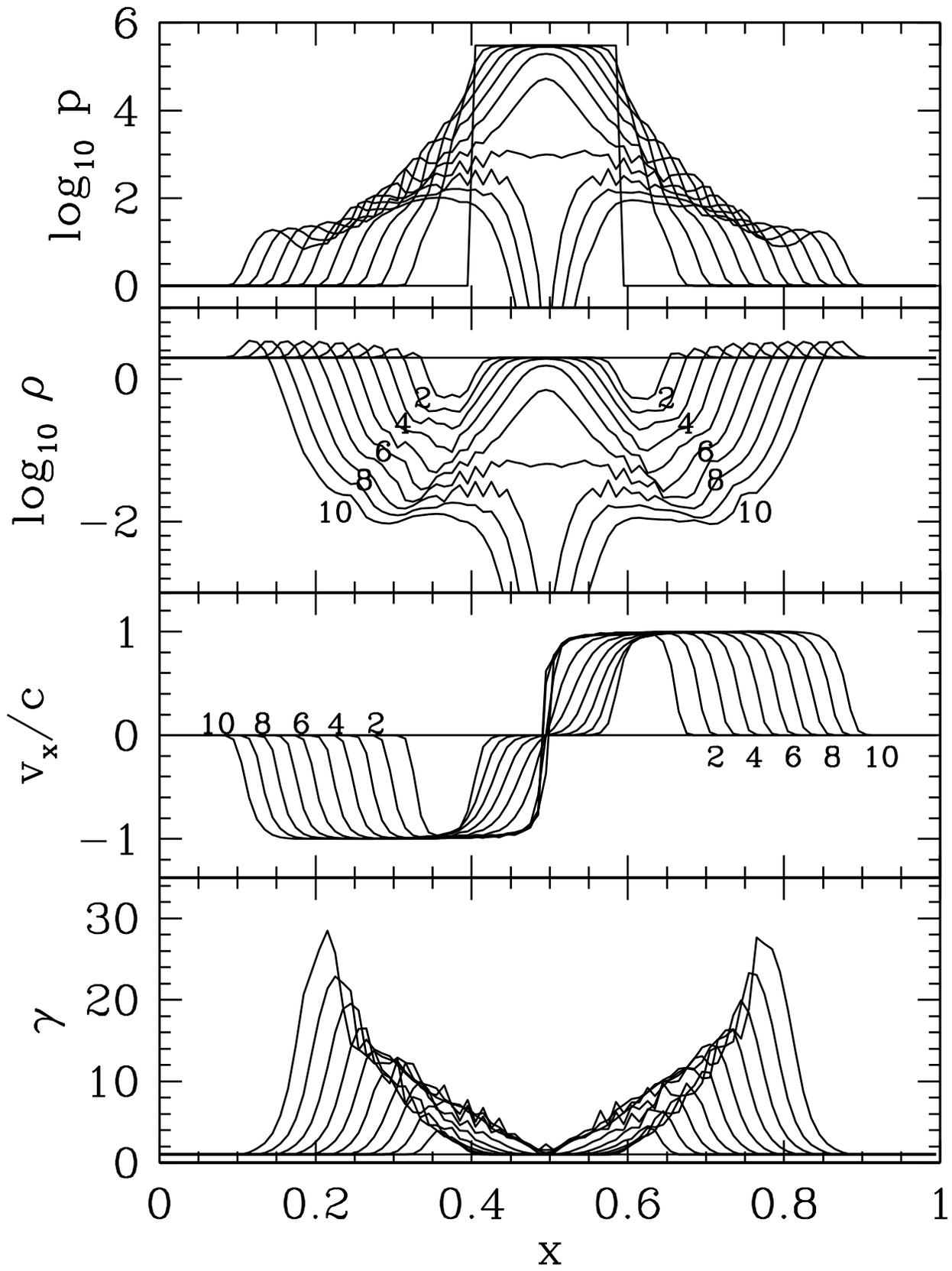
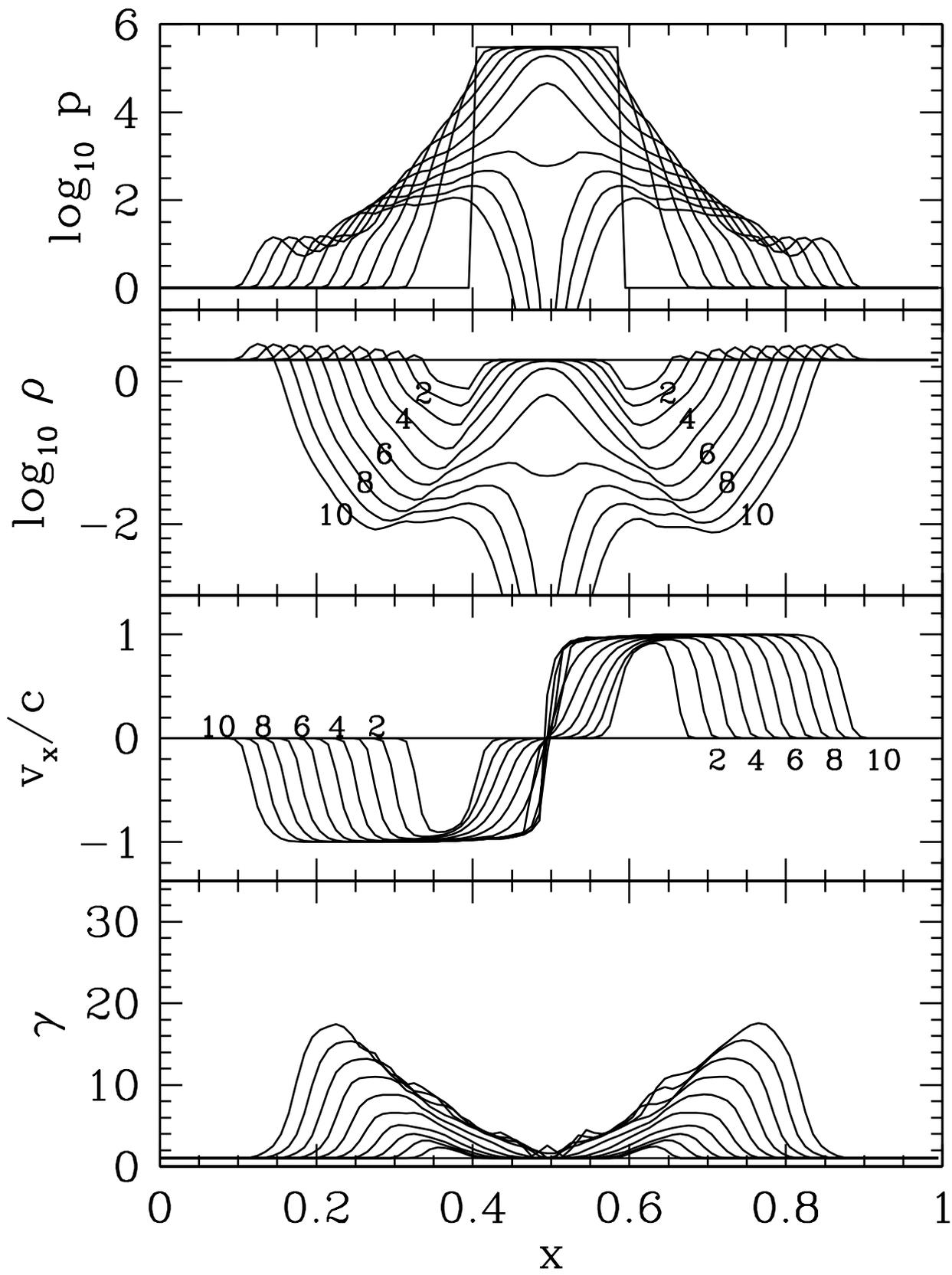
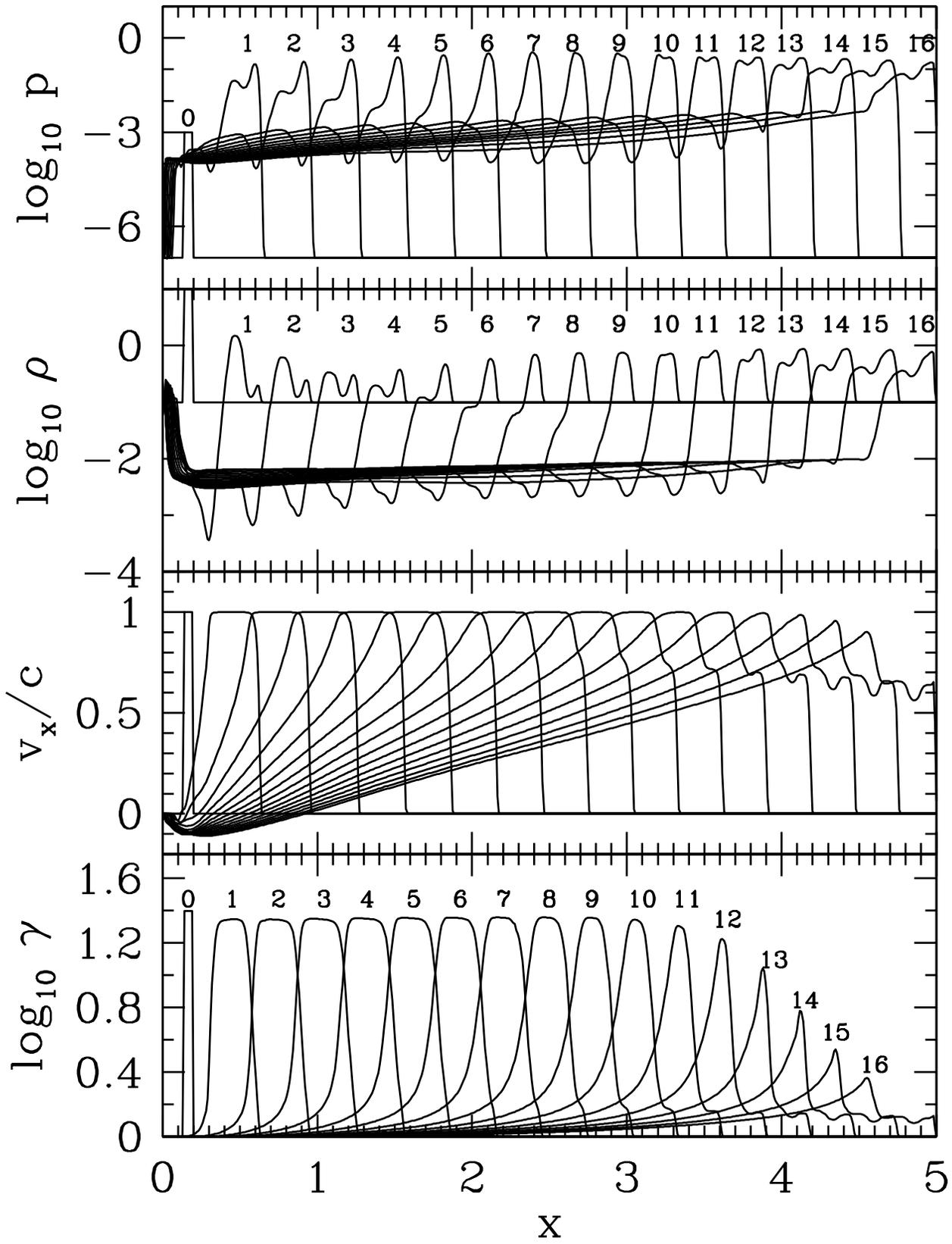
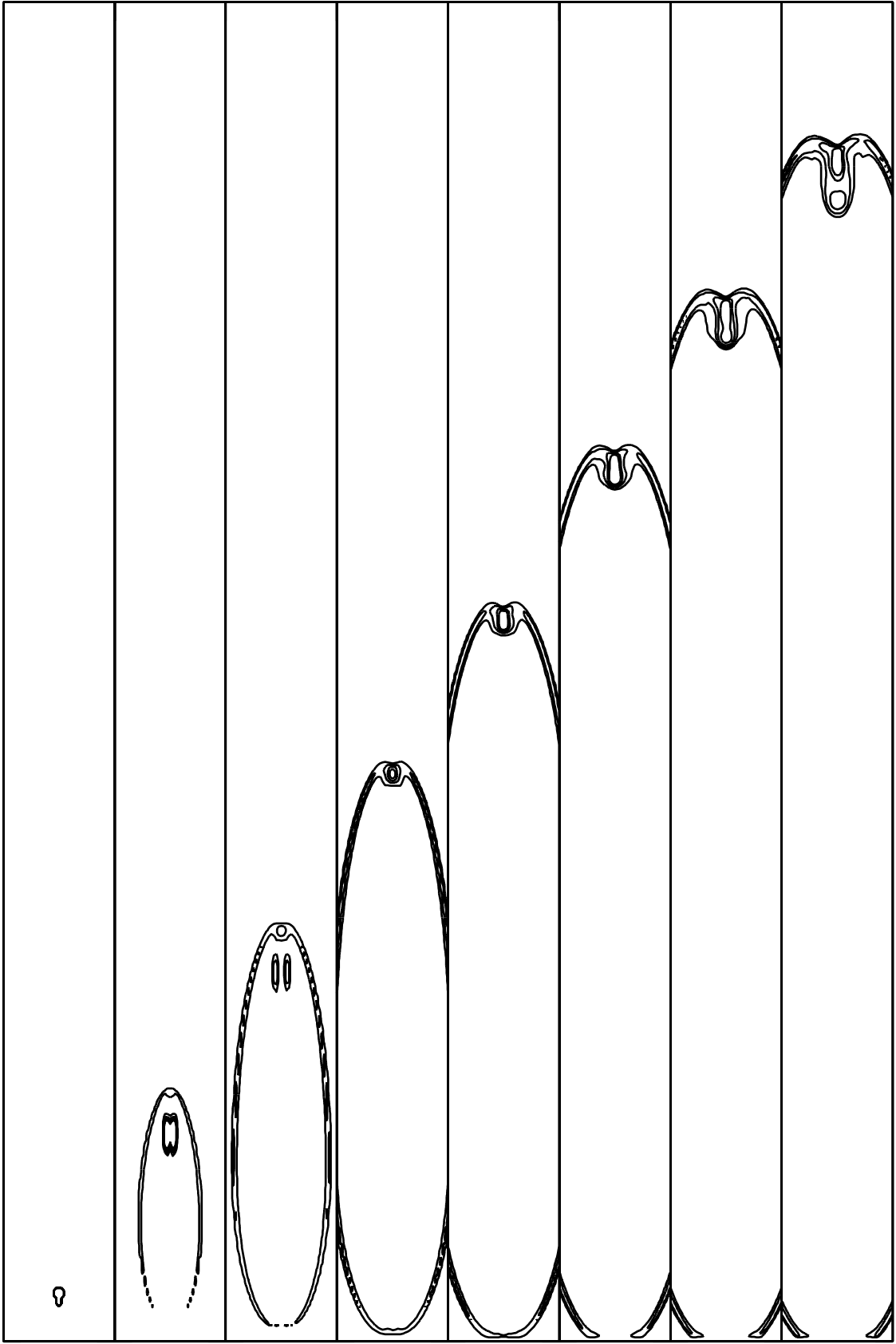
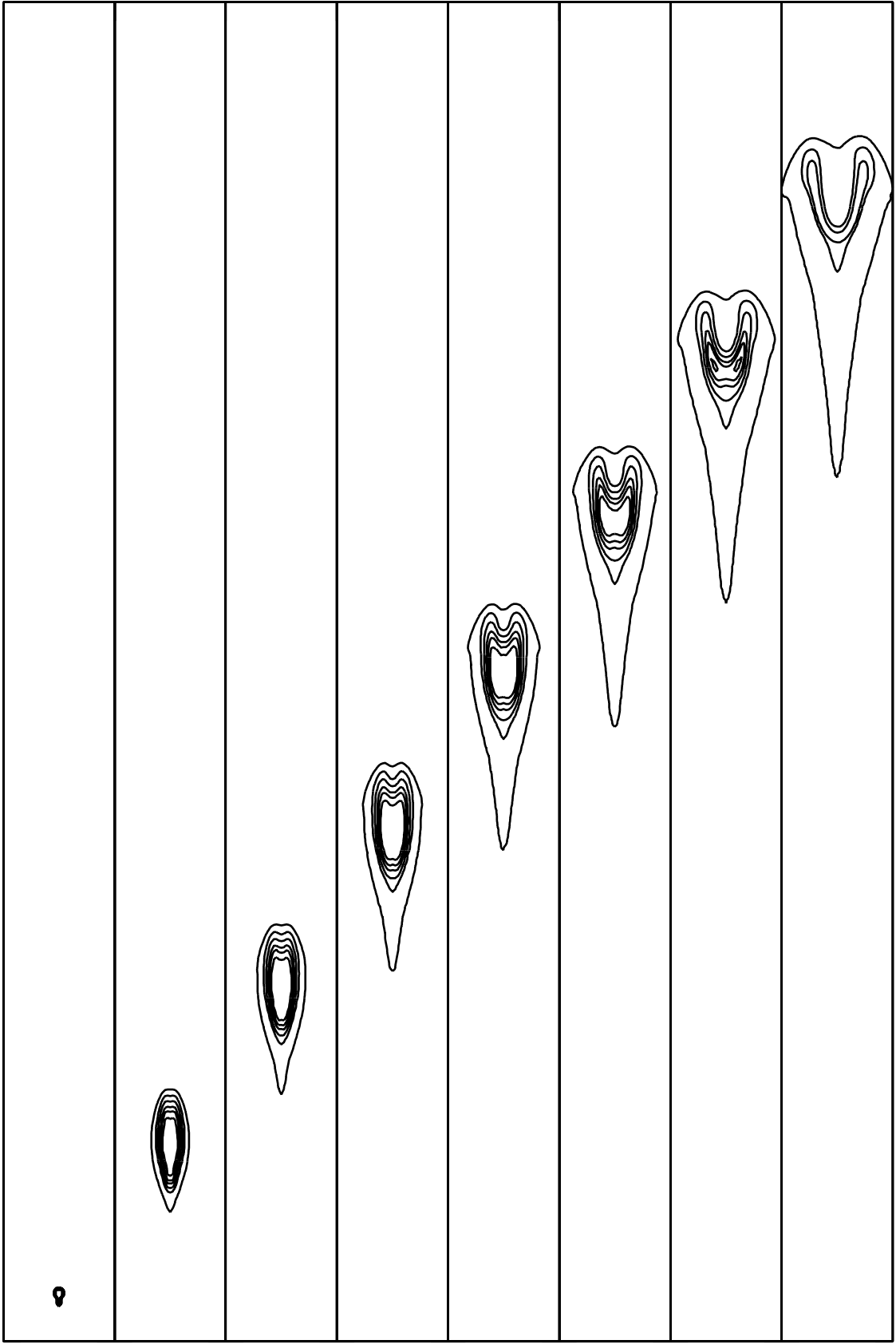
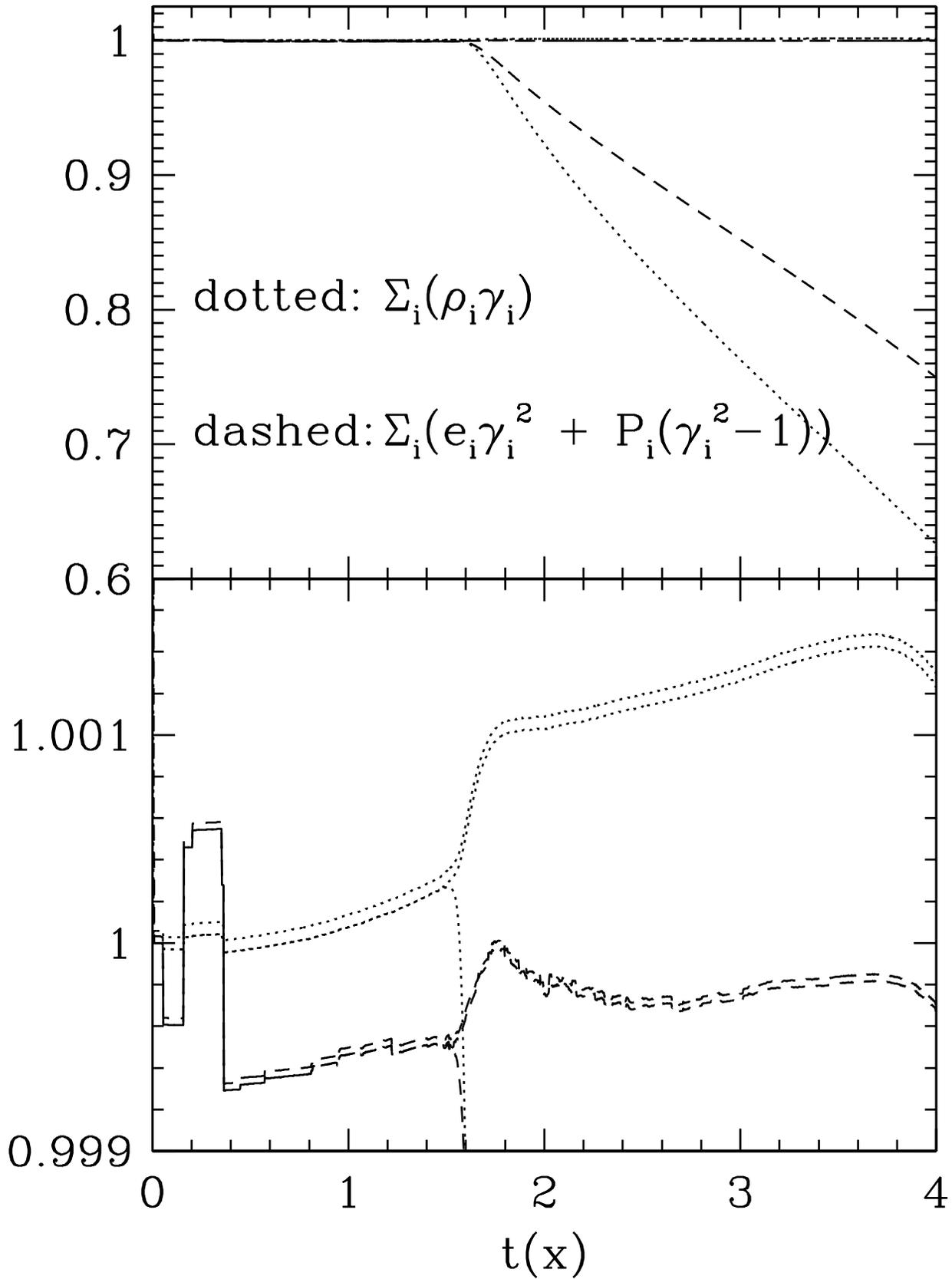
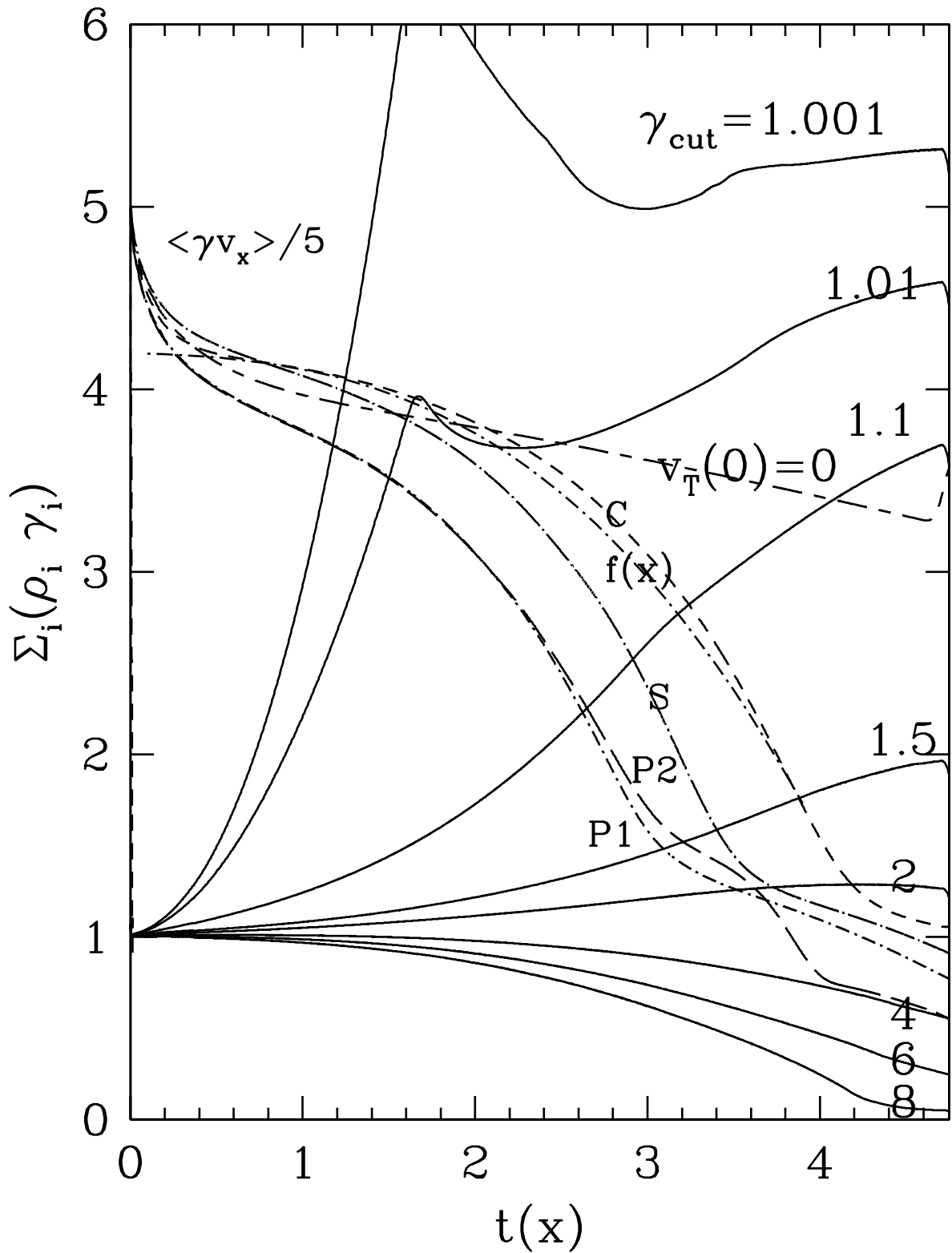
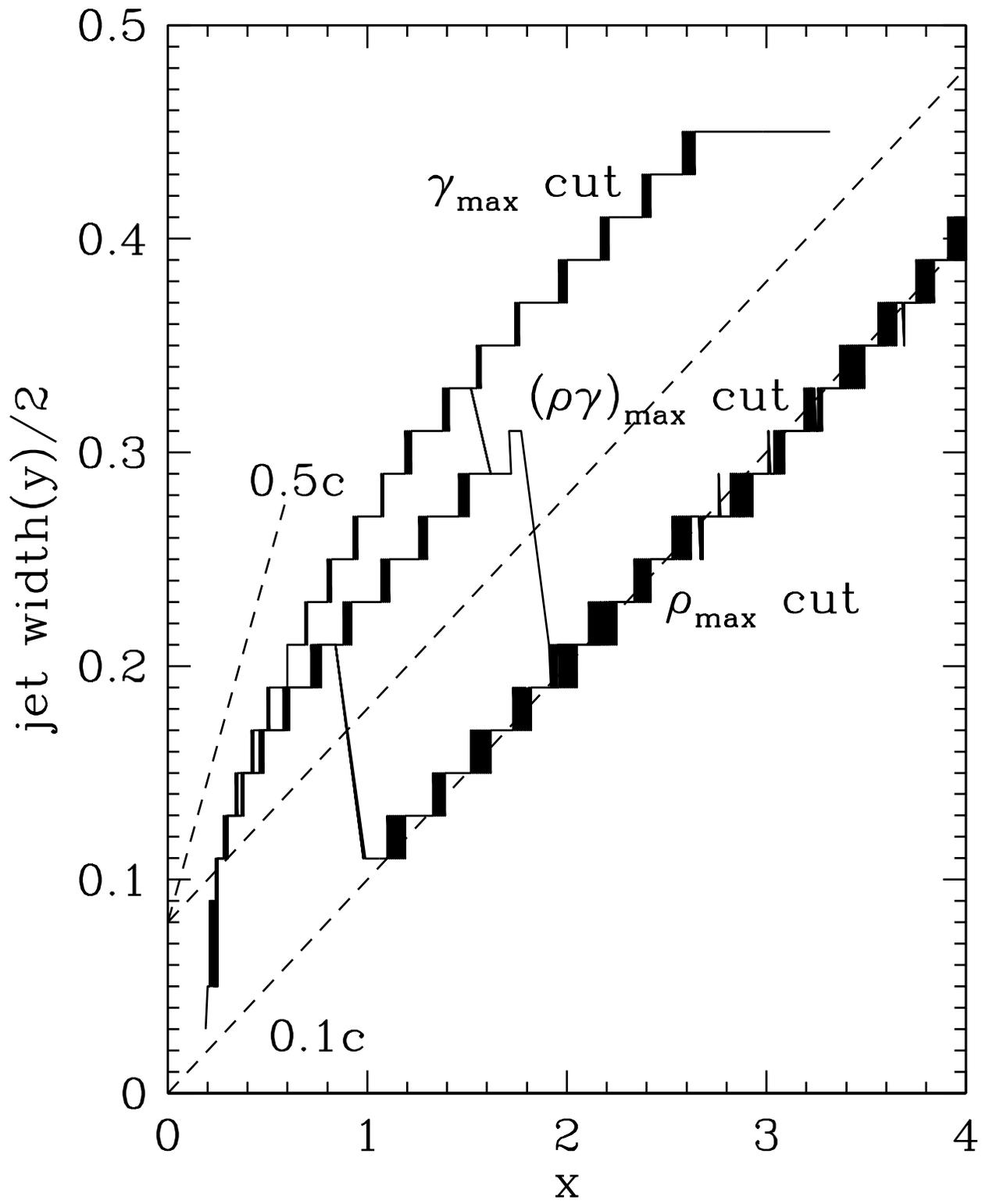
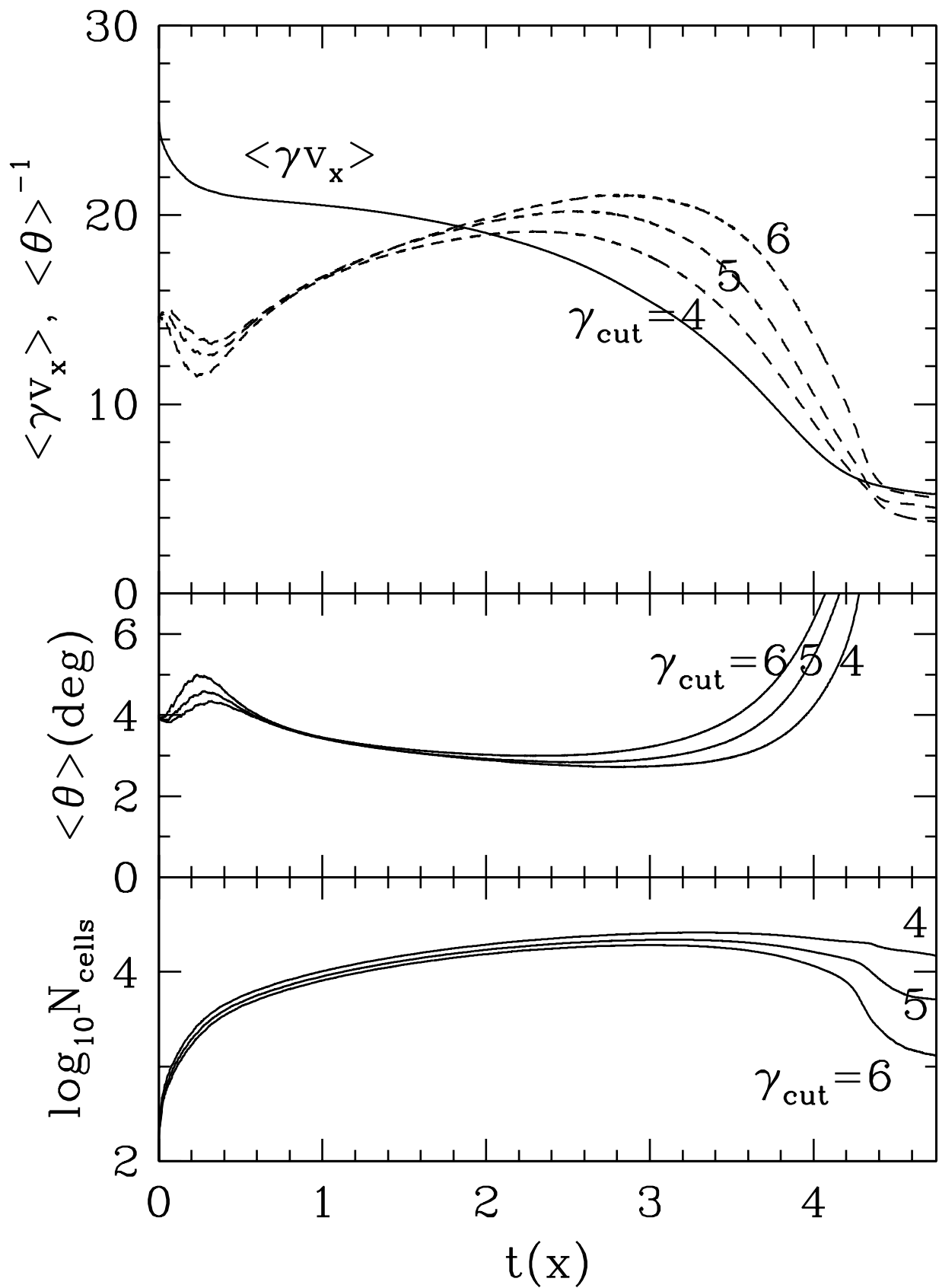
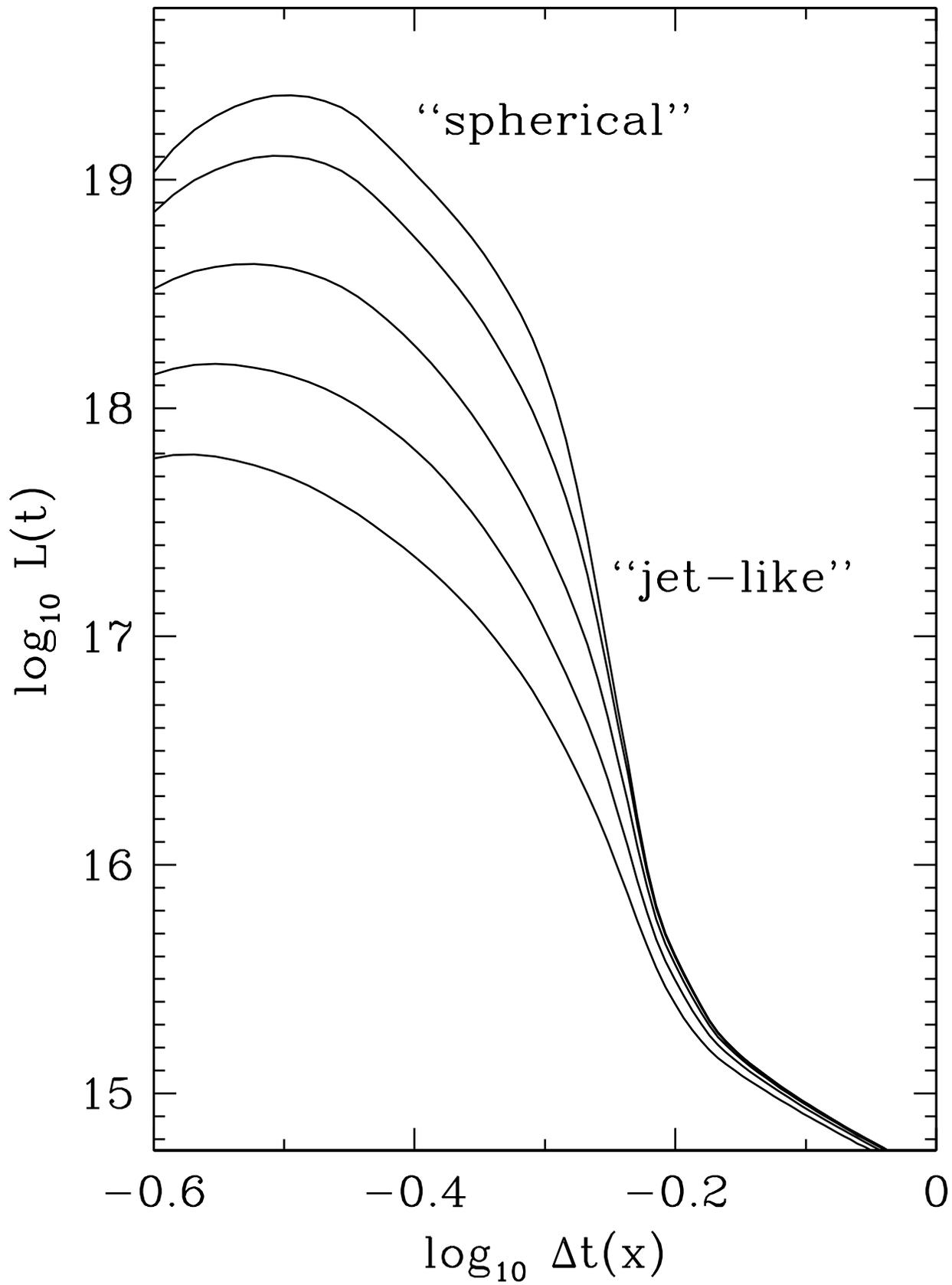
\end{document}